\documentclass[prl,twocolumn,showpacs,amsmath,amssymb]{revtex4}

\usepackage[latin1]{inputenc}
\usepackage{array}
\usepackage{graphicx}

\begin{document}

\title{Local structure of liquid carbon controls diamond nucleation}
\author{L. M. Ghiringhelli$^{1}$\footnote{Max-Planck-Institute for Polymer Research, Ackermannweg 10, 55128 Mainz, Germany}, C. Valeriani$^{2}$\footnote{L.~M. Ghiringhelli and C. Valeriani  contributed equally to this work.}, E. J. Meijer$^1$ and D. Frenkel$^{2}$}
\address{
$^1$ van 't Hoff Institute for Molecular Sciences,Universiteit van
Amsterdam, Nieuwe Achtergracht 166, 1018 WV Amsterdam, The
Netherlands.\\
$^2$ FOM Institute for Atomic and Molecular Physics,Kruislaan 407,
1098 SJ Amsterdam, The Netherlands.
}

\begin{abstract}
Diamonds melt at temperatures above 4000 K. There are no
measurements of the steady-state rate of the reverse process: diamond
nucleation from the melt, because experiments are difficult at
these extreme temperatures and pressures. Using numerical
simulations, we estimate the diamond nucleation rate and find that
it increases by many orders of magnitude when the pressure is
increased at constant supersaturation. The reason is that an
increase in pressure changes the local coordination of carbon
atoms from three-fold to four-fold. It turns out to be much easier
to nucleate diamond in a four-fold coordinated liquid than in a
liquid with three-fold coordination, because in the latter case
the free-energy cost to create a diamond-liquid interface is
higher. We speculate that this mechanism for nucleation control is
relevant for crystallization in many network-forming liquids. On
the basis of our calculations, we conclude that
homogeneous diamond nucleation is likely in carbon-rich
stars and unlikely in gaseous planets.
\end{abstract}

\maketitle

Most liquids can be cooled considerably below their  equilibrium
freezing point before crystals start to form spontaneously in the
bulk. This is caused by the fact that microscopic crystallites are
thermodynamically less stable than  the bulk solid. Spontaneous
crystal growth can only proceed when, due to some rare
fluctuation, one or more micro-crystallites exceed a critical size
(the ``critical nucleus''). An estimate of the rate at which
critical nuclei form in a bulk liquid can be obtained from
Classical Nucleation Theory (CNT)~\cite{Kelton}. This theory
relates $R$, the number of crystal nuclei that form per second per
cubic meter, to $\Delta G_{crit}$, the height of the free-energy
barrier that has to be crossed to nucleate a crystal:
\begin{equation}\label{eqn:rate}
R= \kappa\;  e^{-\Delta G_{crit}/k_{B}T}.
\end{equation}
Here $\kappa$ is a kinetic prefactor, {\it T} is the absolute
temperature and $k_{B}$ is Boltzmann's constant. The nucleation
rate depends strongly on  the height of the nucleation barrier.
CNT predicts the following expression for the height of the
nucleation barrier:
\begin{equation}
\Delta G_{crit} = c \frac{\gamma^{3}_{LS}}{\rho_{S}^{2}|\Delta
\mu|^{2}} , \label{barrier}
\end{equation}
where $\gamma_{LS}$ is the liquid-solid surface free energy per
unit area,  $\Delta \mu$ is the difference in chemical potential
between the solid and the supercooled liquid, and  $\rho_{S}$ is
the number density of the crystalline phase.   The factor $c$
depends on the shape of the nucleus, e.g. $c=16\pi/3$ for a
spherical nucleus. As the nucleation rate depends exponentially on
$\Delta G_{crit}$, a doubling of $\gamma_{LS}$ may change the
nucleation rate by many orders of magnitude. In general, the
kinetic prefactor $\kappa$ in Eq.~\ref{eqn:rate} can be estimated
quite well~\cite{kappa}.

Because of the extreme conditions under which homogeneous diamond
nucleation takes place, there have been no quantitative
experimental studies to determine its rate. Moreover, there exist
no numerical estimates of $\Delta \mu$ and $\gamma_{LS}$ for
diamond in supercooled liquid carbon. Hence, it was thus far
impossible to make even an order-of-magnitude estimate of the rate
of diamond nucleation.

In this Letter, we calculate the diamond nucleation rate {\it R}
in liquid carbon at two state points $\{P$=~85~GPa,
$T$=~5000~K$\}$ and $\{P$=~30~GPa, $T$=~3750~K$\}$ (points $A$ and
$B$ in the carbon phase diagram shown in Fig.~1). At both state
points, the liquid is supercooled by $(T_{m}-T)/T_{m}
\approx$~25~$\%$ below the melting curve of diamond, with $T_{m}$
the melting temperature and $T_{m}^{A}~$=~6600~K and
$T_{m}^{B}$=~5000~K, respectively. Simulations studies of the
diamond melting curve have been reported for pressures up to
400~GPa~\cite{Ghiringhelli05}, 1400~GPa and
2000~GPa\cite{Scandologalli}. The last two studies were carried
out by using ``ab-initio'' Molecular Dynamics. However, it would
be prohibitively expensive to study nucleation using such an
approach. We therefore use a semi-empirical many-body potential
that has been fit to experimentally measured and ab-initio
calculated properties of carbon solid phases and the
liquid~\cite{LCBOP}. We use this model to study diamond nucleation
in a system of 2744 particles in the "low-pressure" (P<100GPa)
region of the phase diagram. In this pressure-range, the
calculated melting line of
Refs.~\cite{Ghiringhelli05,Scandologalli} are in reasonable
agreement. In particular, all the three calculations predict a
melting temperature of about 7000~K at 100~GPa.
\begin{figure}[h!]
\includegraphics[width=1.05\columnwidth,clip]{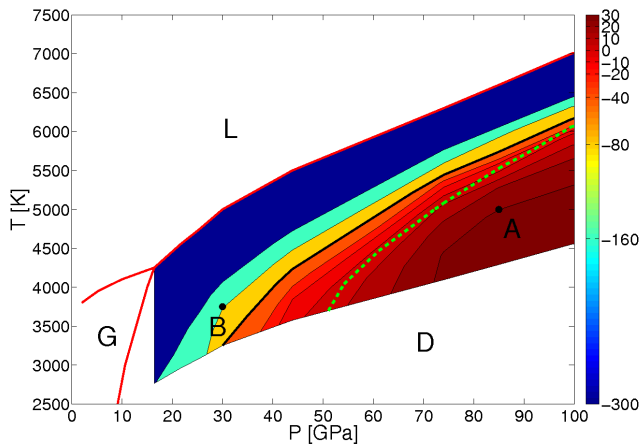}
\caption{The figure shows part of the carbon phase diagram from
Ref.~\cite{Ghiringhelli05} and the iso-nucleation rate zones. The
solid red lines represent the coexistence lines from. $A$ is at $P_{A}$=85~GPa,
$T_{A}$=~5000~K, and $B$ at $P_{B}$=~30~GPa,
$T_{B}$=3750~K. In the text, we give estimates for the nucleation
rate at A and B. Along the green dashed curve, the ratio of
3-fold and 4-fold coordination in the liquid is 1:1. The color
code used in the plot is: numbers on the right indicate the order of magnitude of the nucleation rate
(in $m^{-3}s^{-1}$). The continuous black curve  indicates the boundary of the region where
the nucleation rate is negligible ($<10^{-40} m^{-3}s^{-1}$).}
\label{f1}
\end{figure}

In order to estimate the crystal-nucleation rate, we first
determine the free-energy barrier $\Delta G_{crit}$ to form a
critical nucleus. For state point $A$ we use biased Monte Carlo
simulations (``umbrella sampling''~\cite{umbrella}) to estimate
the excess Gibbs free energy of small crystallites, and find
$\Delta G_{crit,A}$~=~25~$k_{B}T$ for a critical nucleus size of
$N_A=110$~\cite{boxsize}. Knowing $\Delta G_{crit}$ and the kinetic pre-factor
(see Ref.~\cite{kappa}) we estimate the crystal nucleation rate to
be $R_{A}$~=~$10^{30} s^{-1} m^{-3}$. When we tried to follow the
same procedure to compute the diamond nucleation rate at state
point $B$(30~GPa, 3750~K), we failed to reach a critical
nucleus that was small enough to fit in a system of 2744
particles. We therefore had to resort to an indirect way, based on
CNT, to estimate $\Delta G_{crit}$. CNT assumes that $\Delta
G(N)$, the Gibbs free energy difference between a metastable
liquid containing an {\it N}-particle crystal nucleus and a pure
liquid, is given by $\Delta G(N)= S(N)\gamma_{LS}-N|\Delta\mu|$,
where $S(N)$ is the area of the interface between an N-particle
crystallite and the metastable liquid.  In order to determine the
number of particles in a crystallite, we use a spherical-harmonics
based criterion (see e.g. Ref.~\cite{Q6}) that allows us to
distinguish particles in a liquid-like environment from those in a
crystal (diamond or graphite).  The surface area $S(N)$ is given
by $a(N/\rho_{S})^{2/3}$, where the factor $a$ depends on the
geometry of the nucleus.  From our simulations, we can only
determine the product $a\gamma_{LS}$: it is this quantity and the
degree of supersaturation ($\Delta \mu$), that determine the
nucleation rate. In order to calculate $\Delta \mu$, we compute
the temperature dependence of the molar enthalpy difference
($\Delta h$) between the supercooled fluid and the stable crystal
at equal pressures. From $\Delta h$, $\Delta \mu$ is evaluated by
thermodynamic integration from the melting point~\cite{deltamu}.
We find: $|\Delta \mu_{A}/k_BT|$=~0.60 and $|\Delta
\mu_{B}/k_BT|$=~0.77, respectively.

>From the calculated $\Delta G_{crit,A}$ and the number density of
the solid ($\rho_{A}$=~0.191~\AA$ ^{-3}$), we can estimate the
surface free energy per unit area at state point $A$ using
Equation(~\ref{barrier}). Assuming that the critical nucleus is
effectively spherical, we find $\gamma_{LS,A}\approx
0.27$~$k_{B}T$/\AA$^{2}$=1.86~J/m$^{2}$. We stress that, in what
follows, we do not make use of this estimate: rather, we always
employ the combination $a\gamma$ that follows directly from the
simulations.

In state point $B$ we could not follow the same procedure, as a
system of 2744 particles is too small to accommodate a critical
nucleus. In order to estimate $\gamma_{LS,B}$, we therefore
prepared a rod-like crystal in a system with a slab geometry (a
flattened box containing $N\sim$~4000 particles, with lateral
dimensions that are some four times larger than its height). The
crystal rod is oriented perpendicular to the plane of the slab. It
spans the height of the simulation box and is continued
periodically. The cross section of this crystal rod is lozenge
shaped, such that its [111]-faces are in contact with the
liquid~\cite{111}. We used umbrella sampling to determine the
Gibbs free energy of such a crystallite as a function of its size,
both at state points $A$ and $B$. In this way we estimate the
ratio of the surface free energies at $A$ and $B$. We find that $a
\gamma_{LS,B}/ a \gamma_{LS,A} = \gamma_{LS,B}/ \gamma_{LS,A} \sim$~
2.5(the $a$ factors are the same since the shape of the rod
nucleus is approximately the same at the two state points). Since
we know $\gamma_{LS,A}$ from the height of the nucleation barrier
in state point $A$ for a spherical nucleus, we deduce the
corresponding $\gamma_{LS,B}$ for a spherical nucleus. Using our
estimate, $\gamma_{LS,A}\approx$1.86~J/m$^{2}$, we find
$\gamma_{LS,B}\approx$0.68~$k_{B}T/$\AA$^2$= 3.5~J/m$^{2}$. As
$\Delta\mu$ and $\rho_{B}$ are known ($\rho_{B}$=0.17~\AA$^{-3}$),
we can now use CNT to estimate $\Delta G_{crit}$ in state
point $B$. It turns out that, mainly because $\gamma_{LS,B}$ is
2.5 times larger than $\gamma_{LS,A}$, the nucleation barrier in
$B$ is more than ten times higher than in point $A$, thereby
hugely suppressing the nucleation rate ($R_{B} \sim
10^{-80}$~s$^{-1}$m$^{-3}$). We can estimate the size of the
critical (spherical) nucleus at $B$ to be around $N_B=$1700
particles. Thus, one would need a system of at least 17000
particles to contain a critical nucleus and avoid spurious
interactions among its periodic images. Such a system size is
beyond our present computational capacity. In contrast, in the
slab geometry we find that the free energy of a lozenge-shaped
crystal goes through a maximum at a size of $\sim$ 340 particles,
which is much less than the system size (4000 particles).
\begin{figure}[h!]
\includegraphics[width=0.9\columnwidth,clip]{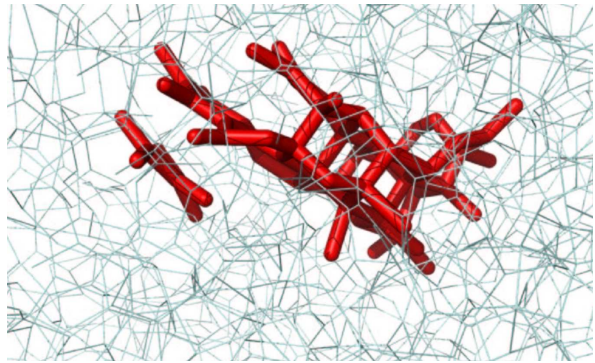}
\caption{Typical snapshot of a small crystalline nucleus of $\sim$~
75 particles obtained at 30~GPa and 3750~K ($B$),
surrounded by mainly three-fold coordinated liquid particles (gray lines). The
nucleus contains both three- (left part) and four-fold (right
part) coordinated particles.} \label{f2}
\end{figure}

To understand the microscopic origin for the large difference in
nucleation rates in state points $A$ and $B$, it is useful to
compare the local structure of the liquid phase in both state
points.  It turns out that the liquid structure in state points
$A$ and $B$ is markedly different (see also Refs.
~\cite{Ghiringhellidlike,Ghiringhellijocm}): liquid carbon is
mainly four-fold coordinated at state point $A$ ($20\%$ three-fold
and $80\%$ four-fold), while at the lower temperatures and
pressures of point $B$, the coordination in the liquid resembles
that of the graphite and is mainly three-fold coordinated ( $5\%$
two-fold, $85\%$ three-fold and $10\%$ four-fold ). Apparently, it
is less favorable to create an interface between a diamond and a
graphitic liquid than between a diamond and a four-fold
coordinated liquid. The destabilizing effect of the graphitic
liquid on the diamond nuclei is most pronounced for small nuclei
(large surface-to-volume ratio). In fact, in state point $B$,
nuclei containing less than $25$ particles tend to be graphitic in
structure, with a small number of four-fold coordinated particles
linking the different graphite planes. Nuclei containing up to
$60$ particles show a mixed graphite-diamond structure, whereas
larger nuclei have a diamond bulk like structure, but the surface
remains graphitic in nature (see Fig.~2). The unusual surface
structure of the diamond nucleus is an indication of the poor
match between a diamond lattice and a three-fold coordinated
liquid.

There are many network-forming liquids that, upon changing
pressure and temperature, undergo profound structural changes or
even liquid-liquid phase transitions~\cite{liqliq}. Our
simulations on carbon indicate that such a change in the local
coordination in the liquid has dramatic consequences for the rate
of crystal nucleation. Experiments on a completely different class
of materials, viz. liquid metals~\cite{kelton2}, suggest that the
local structure, in particular, local icosahedral packing,  may
interfere with direct nucleation of crystals. What is interesting
about the present simulations is that we show  that the ease of
homogeneous crystal nucleation from one-and-the-same meta-stable
liquid can be tuned by changing its pressure, and thereby its
local structure.

The thermodynamic conditions we discuss are relevant for
experiments that study nucleation in compressed, laser-melted
carbon. In addition, homogeneous nucleation of diamond may have
taken place in carbon-rich white dwarf~\cite{whitedwarf}. It has
also been suggested that diamonds could also have formed in the
carbon-rich middle layer of Uranus and Neptune~\cite{uranus}. The
present work allows us to make a rough estimate of the conditions
that are necessary to yield appreciable diamond nucleation on
astronomical timescales.

Neither white dwarfs nor planets consist of pure carbon.
Nevertheless, it is useful to estimate an upper bound to the
diamond nucleation rate by considering the rate at which diamonds
would form in a hypothetical environment of pure carbon. To this
end we use our numerical data on the chemical potential of liquid
carbon and diamond and our numerical estimate of the
diamond-liquid surface free energy, to estimate the nucleation
barrier of diamond as a function of temperature and pressure. We
then use CNT to estimate the rate of diamond nucleation~\cite{interp}. The results are shown in Fig.~1. The figure
shows that there is a region of some 1000~K below the freezing
curve (continuous red line) where diamond nucleation is less than
$10^{-40} m^{-3}s^{-1}$. If the rate is lower than this
number, not a single diamond could have nucleated in a
Uranus-sized body during the life of the universe. As can be seen
from the figure, our simulations for state point B are outside the
regime where observable nucleation would be expected. However,
this figure provides just an upper bound to the rate of diamond
nucleation. In practice, the carbon concentration is somewhat less
in carbon-rich stars ($\sim$50\%)~\cite{whitedwarf}, and much less
(1-2\%~\cite{uranus}) in Uranus and Neptune.

In Fig.~3, we show the effect of dilution on the regime where
diamond nucleation is possible.
\begin{figure}[h!]
\includegraphics[width=1.05\columnwidth,clip]{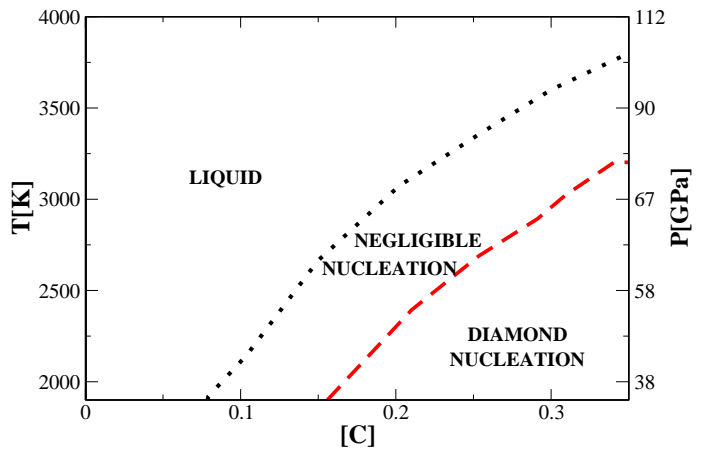}
\caption{Diamond nucleation boundaries as a function of carbon
concentration: the rate is zero (no thermodynamic driving force to nucleation) in the top region (liquid),
it is negligible ($<10^{-40} m^{-3}s^{-1}$) in the middle region and
non-negligible ($>10^{-40} m^{-3}s^{-1}$) in the bottom-right
region. The nucleation rate is negligible when it corresponds
to less than one nucleus per Uranus-sized planet over a period of
10$^{10}$  years. The left hand y-axis represents the temperature;
the right-hand y-axis indicates the corresponding pressure for a
Uranus-like isentrope (see Ref.~\cite{uranus}).} \label{f3}
\end{figure}
To simplify this figure, we do not vary pressure and
temperature independently but assume that they follow the
adiabatic relation that is supposed to hold along the isentrope of
Uranus~\cite{uranuscurve}. We make the assumption that nucleation
takes place from an ideal mixture~\cite{idmix} of C, N,O and
H~\cite{HansenBarrat}. Not surprisingly, Fig.~3 shows that
dilution of the liquid decreases the driving force for
crystallization to the extent that no diamond phase is expected
for $C$ concentrations of less than 8\%. As before, there is a
wide range of conditions where diamonds could form in principle,
but never will in practice. Assuming that, for a given pressure,
the width of this region is the same as in the pure $C$ case
(almost certainly a serious underestimate), we arrive at the
estimate in Fig.~3 of the region where nucleation is
negligible (i.e. less than one diamond per planet per
life-of-the-universe). From this figure, we see that quite high
Carbon concentrations (over 15\%) are needed to get homogeneous
diamond nucleation. Such conditions do exist in white dwarfs, but
certainly not in Uranus or Neptune. Hence, our simulations
indicate that it is extremely unlikely that diamonds could ever
have nucleated from the carbon-rich middle layer of Uranus and
Neptune.

We gratefully thank D.~J. Stevenson, E.~Tosatti, S.~Iacopini, E.~Sanz, P.~R.~ten~Wolde and H.~L.~Tepper for helpful suggestions, and
gratefully acknowledge J.~Los and A.~Fasolino, whose MC
code~\cite{LCBOP} was adapted for our calculations.
The work of the FOM Institute is part of the
research program of FOM and is made possible by financial support
from the Netherlands Organization for Scientific Research (NWO),
partly through grant 047.016.001.


\bibliography{carbon_3}

\begin{thebibliography}{20}
\bibitem{kappa}
The CNT expression for the $\kappa$ is:
$\kappa$=$Z \rho_{L} 24 D_{S} n_{crit}^{2/3} / \lambda^{2}$
where $\rho_{L}$ is the liquid number density,
$D_{S}$ the liquid self-diffusion coefficient, $\lambda$ a
typical atomic jump distance, and $n_{crit}$ the number of
particles belonging to the critical nucleus.
$Z$=$\sqrt{\frac{|\Delta \mu|}{6 \pi k_{B}T n_{crit}}}$ is
Zeldovitch factor~\cite{Kelton}.
\bibitem{boxsize}
To determine the critical nucleus we must locate the top of the
free energy barrier. For high barriers, the spontaneous formation
of a critical nucleus is a rare event. We therefore use biased MC
sampling to compute the probability that a critical nucleus forms.
>From this probability we can then directly derive the free energy
of this nucleus. In order to suppress finite-size effects, we
choose a simulation-box diameter $L$ that is at least 4 times
larger than the radius of the crystal nucleus. A spherical nucleus
with a radius of 0.25 $L$ occupies $\sim$7\% of the box volume and
contains about 10\% of the total number of particles.
\bibitem{Q6}
We used a local order parameter as introduced in Ref.~\cite{umbrella};
rather than the 6th order harmonic, we
found that a 3rd order harmonic can capture the
correct symmetry for both graphite and diamond lattices.
\bibitem{deltamu}
In detail:
$\Delta (\beta \mu) =  \int_{\beta_{\mathrm{coex}}}^{\beta_{1}} {
\left< [h^{S}(\beta) - h^{L}(\beta)] \right>_{P,\beta}} \ d\beta$
where $h^{S}(\beta)$ ($h^{L}(\beta$) is the specific enthalpy of
the solid (liquid) and the integration path is an isobar from the
coexistence temperature ($\beta_{\mathrm{coex}}=1/(k_B
T_\mathrm{coex}))$ to the temperature of the selected state point.
\bibitem{111}
The [111] planes are the most stable for the diamond lattice.
Note that when graphite and diamond structures compete at 30~GPa
(see Fig.~2), the [001] graphite sheets transform into [111]
diamond planes. We can write the size ratio of  lozenge-shaped and
spherical critical nuclei at the same state point as:
$ N_{\mathrm{crit}}^{\mathrm{sph}}/
N_{\mathrm{crit}}^{\mathrm{loz}} = 8 \pi \gamma \sin \theta /(3
\rho_{s} \Delta \mu h)$
where $\gamma$ is the lozenge-shaped nucleus interfacial free energy,
$\theta$ the angle between two adjacent side (70.05$^\circ$), $\rho_{s}$
 the solid density (0.17~\AA$ ^{-3}$) and
$\beta \ |\Delta \mu|$=0.77 in $B$, and $h$ the slab height (10 \AA).
Our numerical results indicate that a
spherical critical nucleus is about 5 times
larger than a rod-like nucleus in the (unphysical)
periodic slab geometry.
\bibitem{interp}
In view of the strong dependence of $\gamma_{LS}$ on the atomic
coordination in the liquid phase,  we assume that
$\gamma_{LS}(P,T)$ is linear with $c_4$, the equilibrium
concentration of 4-fold coordinated atoms at the selected state
point.
\bibitem{idmix}
We extend our results by making use of: $\beta \ \Delta \mu =
\beta \ \Delta \mu_{0} + \beta \ \ln[C]$, where $\beta \ \Delta
\mu_{0}$ is the chemical potential of the pure carbon and $[C]$ is
the concentration in the mixture.
\end{thebibliography}

\end{document}